\documentclass[twocolumn,osajnl,preprint,showpacs,10pt]{revtex4}  

\usepackage{graphicx}
\usepackage{dcolumn}
\usepackage{bm}

\newcommand{\ud}{d}
\newcommand{\eqn}[1]{(\ref{#1})}
\newcommand{\fig}[1]{Fig. \ref{#1}}
\newcommand{\figs}[1]{Figs. \ref{#1}}

\begin{document}
%
\title{Non-Gaussian Statistics of the Soliton Timing Jitter due to Amplifier Noise}
\author{Keang-Po Ho}
\affiliation{StrataLight Communications, Campbell, CA 95008.}
\email{kpho@stratalight.com}

\date{\today}%
%
\begin{abstract}
Based on the first-order perturbation theory of soliton, the Gordon-Haus timing jitter induced by amplifier noise is found to be non-Gaussian distributed.
Compared with Gaussian distribution given by the linearized perturbation theory, both frequency and timing jitter have larger tail probability.
The timing jitter has a larger discrepancy to Gaussian distribution than that of frequency jitter.
\end{abstract}



\maketitle

The Gordon-Haus timing jitter in fiber soliton due to amplifier noise is usually assumed to be Gaussian distributed \cite{gordon86,  moore03, iannone} when the first-order perturbation theory of soliton \cite{kivshar89, kaup90, georges95} is used.
Previous works showed that the non-Gaussian timing jitter is induced by soliton interactions \cite{menyuk95, georges95a, georges96} and regeneration \cite{leclerc99, falkovich01} but not by amplifier noise alone.
When the first-order perturbation of soliton is linearized \cite{gordon86, iannone}, the Gordon-Haus timing jitter is indeed Gaussian distributed.
However, if the ordinary differential equations from the first-order perturbation is not linearized, as shown later, the amplitude, frequency and timing jitters are all non-Gaussian distributed.

In this letter, the amplitude jitter is found to be noncentral chi-square ($\chi^2$) distributed, confirming the previous simulation results \cite{moore03} and the non-Gaussian statistics \cite{falkovich01}.
Both the frequency and timing jitters are also non-Gaussian distributed.

From the first-order perturbation theory of soliton \cite{kivshar89, kaup90, georges95}, with amplifier noise alone, the soliton parameters are evolved according to the following ordinary (or stochastic) differential equations:

\begin{eqnarray}
\frac{d A}{d \zeta} & = & n_A(\zeta), \label{amp}\\
\frac{d \Omega}{d \zeta} & = & n_{\Omega}(\zeta), \label{freq} \\
\frac{d T}{d \zeta} & = &- \Omega + n_T(\zeta). \label{time}
\end{eqnarray}

\noindent Not used here, the phase perturbation is not shown in \eqn{amp}-\eqn{time}. 
All noise terms of $n_A(\zeta), n_\Omega(\zeta)$, and $n_T(\zeta)$ are independent Gaussian processes with autocorrelation of \cite{georges95, georges96} 

\begin{eqnarray}
E\{n_A(\zeta_1) n_A(\zeta_2)\} & = &A \sigma_n^2 \delta(\zeta_1 - \zeta_2), \label{ampvar} \\
E\{n_\Omega(\zeta_1) n_\Omega(\zeta_2)\} &=& \frac{A}{3} \sigma_n^2 \delta(\zeta_1 - \zeta_2), \label{freqvar} \\
E\{n_T(\zeta_1) n_T(\zeta_2)\} &=& \frac{\pi^2}{12A^3} \sigma_n^2 \delta(\zeta_1 - \zeta_2), \label{timevar} 
\end{eqnarray}

\noindent where $\sigma_n^2$ is the noise variance and $E\{\cdot\}$ denotes expectation.
From \eqn{ampvar} to \eqn{timevar}, the variances depend on the amplitude of $A(\zeta)$ and amplitude jitter transfers to both frequency and timing jitter.
If the amplitude in the variances of \eqn{ampvar} to \eqn{timevar} is assumed to be constant [$A(\zeta) \approx A(0) = A_0$] as a first-order approximation, amplitude, frequency and timing jitters are indeed Gaussian distributed.

The inclusion of amplitude jitter in \eqn{ampvar} to \eqn{timevar} is still within the equations of the first-order soliton perturbation theory \cite{kivshar89,kaup90, georges95}. 
The nonlinear first-order perturbation can be interpreted the repeated usage of the linearized first-order perturbation \cite{iannone}.

Based on \eqn{amp} and \eqn{ampvar}, the stochastic differential equation (SDE) of the amplitude jitter is $d A = \sqrt{A} d w_A$ where $w_A$ is a Wiener process with autocorrelation of $E\{w_A(\zeta_1) w_A(\zeta_2)\} = \sigma_n^2 \min(\zeta_1, \zeta_2)$.
With an initial value of $A(0) = A_0$, the amplitude jitter is

\begin{equation}
A(\zeta) = \left( A_0^{1/2} +\frac{w_A}{2} \right)^2
\label{ampsol}
\end{equation}

\noindent as a noncentral $\chi^2$-distributed random process \cite{humblet91, shum96, moore03} with variance parameter of $\frac{1}{4}\sigma_n^2 \zeta$.
The amplitude jitter of \eqn{ampsol} is the solution of a Stranovich but not Ito SDE \cite{gardiner2}.

Using \eqn{freq} and \eqn{freqvar}, the frequency jitter is

\begin{equation}
\Omega(\zeta) = \int_0^\zeta \left( A_0^{1/2} +\frac{w_A(\zeta_1)}{2} \right) \ud w_\Omega,
\label{Fjit}
\end{equation}

\noindent where $w_\Omega$ is a Wiener process with autocorrelation of $E\{w_\Omega(\zeta_1) w_\Omega(\zeta_2)\} = \frac{1}{3} \sigma_n^2 \min(\zeta_1, \zeta_2)$ and independent of the Wiener process of $w_A$.

The timing jitter equation of \eqn{time} has two terms, the first-term of $-\Omega$ gives the Gordon-Haus timing jitter increasing with $\zeta^3$ and second term of $n_T(\zeta)$ is just the projection of amplifier noise into timing jitter \cite{gordon86}.
The first term of Gordon-Haus timing jitter is far more interesting than the second term.
The SDE of $d T_{\mathrm{GH}} = - \Omega d \zeta$ has a solution of

\begin{equation}
T_{\mathrm{GH}}(\zeta) = - \int_0^\zeta (\zeta - \zeta_1) \left( A_0^{1/2} +\frac{w_A(\zeta_1)}{2} \right) \ud w_\Omega.
\label{Tjit}
\end{equation}

Similar to option pricing with stochastic volatility \cite{stein91}, the characteristic functions of the frequency $\Omega(\zeta)$ and Gordon-Haus timing jitter $T_{\mathrm{GH}}(\zeta)$ are

\begin{eqnarray}
\Psi_{\Omega(\zeta)}(\nu) = G_{1}\left(\frac{\nu^2 \sigma^2_n}{6}\right),\label{cfFreq} \\
\Psi_{T_{\mathrm{GH}}(\zeta)}(\nu) = G_{2}\left(\frac{\nu^2 \sigma^2_n}{6}\right), \label{cfTime}
\end{eqnarray}

\noindent and

\begin{eqnarray}
G_1(\lambda) &=& E\left\{e^{-\lambda \int_0^\zeta \left( A_0^{1/2} +\frac{w_A(\zeta_1)}{2} \right)^2 \ud \zeta_1} \right\}, \label{cfC1}\\
G_2(\lambda) &=& E\left\{e^{-\lambda \int_0^\zeta (\zeta - \zeta_1)^2\left( A_0^{1/2} +\frac{w_A(\zeta_1)}{2} \right)^2 \ud \zeta_1} \label{cfC2}\right\} 
\end{eqnarray}

\noindent where $G_1(-\lambda)$ and $G_2(-\lambda)$ are the moment generating functions of $\int_0^\zeta \left( A_0^{1/2} + \frac{1}{2} w_A \right)^2 \ud \zeta_1$ and $\int_0^\zeta (\zeta - \zeta_1)^2 \left( A_0^{1/2} + \frac{1}{2}w_A \right)^2 \ud \zeta_1$, respectively.

Based on the Cameron-Martin integral \cite{cameron45}, we get

\begin{equation}
G_1(\lambda) = \frac{
          \exp \left[- \frac{2A_0}{\sigma_n} \sqrt{\lambda} \mathrm{tanh} 
           \left(\frac{\zeta \sigma_n}{2} \sqrt{\lambda} \right)  \right]}
{ \mathrm{cosh}^{1/2}\left(\frac{\zeta \sigma_n}{2} \sqrt{\lambda} \right)}
\label{cfC1a}
\end{equation}

\noindent and

\begin{equation}
G_2(\lambda) = \left(f_\lambda(\zeta) \over f_\lambda(0)\right)^{1/2}
\exp(\lambda^2 \sigma^2_n A_0 \beta^2 - \lambda A_0 \zeta^3/3),
\label{cfC2a}
\end{equation}

\noindent where

\begin{equation}
\beta^2 = \int_0^\zeta \left[
  \frac{1}{f_\lambda(\zeta_1)}
    \int_{\zeta_1}^{\zeta} (\zeta - \zeta_2)^2 f_\lambda(\zeta_2) \ud \zeta_2 \right]^2 \ud \zeta_1,
\end{equation}

\noindent and

\begin{equation}
f_\lambda(\zeta_1) = \sqrt{\zeta - \zeta_1} 
        I_{-\frac{1}{4}}\!\!\left[ \frac{\sqrt{\lambda} \sigma_n}{2} (\zeta - \zeta_1)^2 \right],
\end{equation}

\noindent where $I_{\nu}(\cdot)$ is the $\nu^{\mathrm{th}}$-order modified Bessel function of the first kind.
The function of $G_1(\lambda)$ \eqn{cfC1a} is similar to the characteristic function of nonlinear phase noise \cite{ho03sta, ho03asy}.
Using $\int_0^1 x^{\nu+1}I_\nu(ax) \ud x = I_{\nu+1}(a)/a$ \cite{table}, 
we get

\begin{equation}
 \int_{\zeta_1}^{\zeta} (\zeta - \zeta_2)^2 f_\lambda(\zeta_2) \ud \zeta_2
= \frac{(\zeta - \zeta_1)^{3/2} }{ \sqrt{\lambda} \sigma_n}
     I_{\frac{3}{4}}\!\!\left(\frac{\sqrt{\lambda} \sigma_n}{2} (\zeta - \zeta_1)^2\!\! \right),
\end{equation}

\noindent and

\begin{equation}
\beta^2 = \sqrt{2}(\sqrt{\lambda} \sigma_n)^{-7/2} \int_0^{\sqrt{\lambda} \sigma_n \zeta^2/2} 
  x^{1/2}  I^2_{\frac{3}{4}}(x)I^{-2}_{-\frac{1}{4}}(x) \ud x.  
\end{equation}

If the amplitude jitter is approximated as $A(\zeta) \approx A_0$ in both variances of \eqn{ampvar} and \eqn{freqvar}, we get

\begin{equation}
G_1(\lambda) \approx \exp(- \lambda A_0 \zeta), G_2(\lambda) \approx \exp\left( - \frac{1}{3}\lambda A_0 \zeta^3\right),  
\label{cfApprox}
\end{equation}

\noindent that are valid for $A_0 \gg \sigma_n^2 \zeta$. 
With the approximation of \eqn{cfApprox}, the characteristic functions of \eqn{cfFreq} and \eqn{cfTime} are zero-mean Gaussian characteristic functions with variance of 

\begin{equation}
\sigma_\Omega^2(\zeta) = \frac{1}{3} A_0 \sigma_n^2 \zeta \mbox{~~and~~} 
\sigma_{T}^2(\zeta) = \frac{1}{9} A_0 \sigma_n^2 \zeta^3,
\label{sigma}
\end{equation}

\noindent  respectively. 
Note that the timing jitter has a variance increase with $\zeta^3$ \cite{gordon86}.

From the frequency jitter of \eqn{Fjit}, the non-Gaussian distribution is induced by the term of $\frac{1}{2} \int_0^\zeta w_A \ud w_\Omega$, i.e., the noise and noise interaction.
The second-order soliton perturbation also includes noise and noise interaction \cite{kaup91, haus97}.
However, the equations of \eqn{amp} to \eqn{time} with noise variances of \eqn{ampvar} to \eqn{freqvar} are directly from the first-order perturbation of soliton \cite{kivshar89, kaup90, georges95}. 
Similarly, the non-Gaussian timing jitter of \eqn{Tjit} is induced by the term of $\frac{1}{2} \int_0^\zeta (\zeta - \zeta_1) w_A \ud w_\Omega$ that also includes noise-noise interaction.

\begin{figure}
\centerline{
  \begin{tabular}{cc}
	 \includegraphics[width = 0.25 \textwidth]{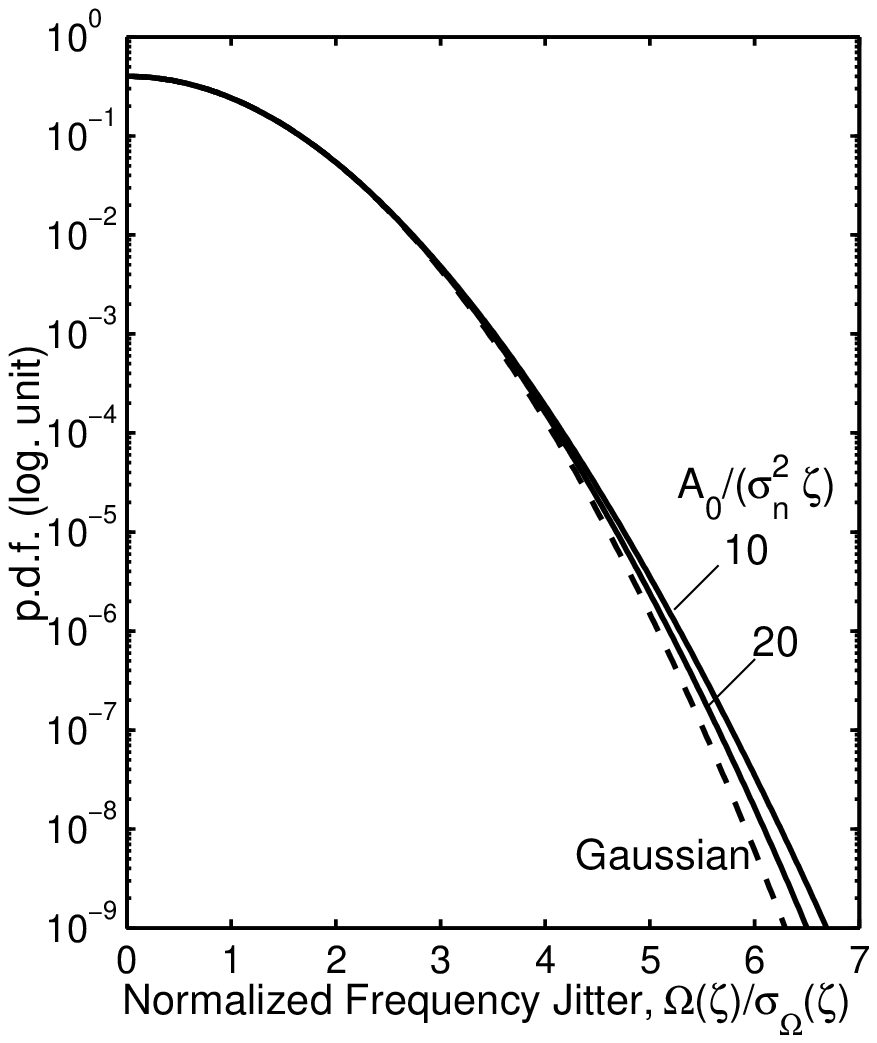} & \includegraphics[width = 0.25 \textwidth]{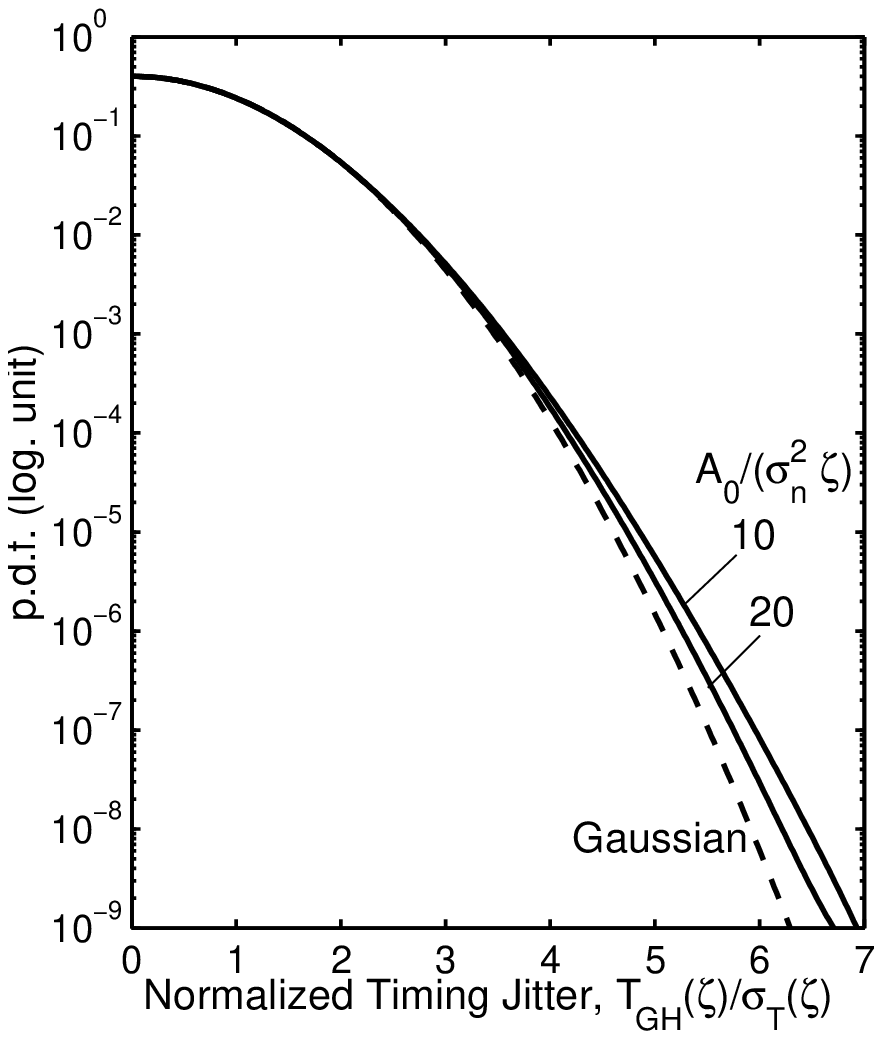} \\
	 (a) &  (b) 
  \end{tabular}
}
\caption{The probability density functions of (a) frequency and (b) timing jitter as compared with Gaussian distribution.}
\label{figpdf}
\end{figure}

The probability density functions of frequency and Gordon-Haus timing jitter are the inverse Fourier transforms of the corresponding characteristic functions of \eqn{cfFreq} and \eqn{cfTime}, respectively. 
\figs{figpdf} plot the probability density of frequency and Gordon-Haus timing jitter as compared with the Gaussian distribution. 
The probability density functions are shown for signal-to-noise ratio (SNR) of $A_0/(\sigma_n^2 \zeta) = 10$ and ${20}$.
The horizontal axis is normalized with respect to the standard deviation of frequency $\sigma_\Omega(\zeta)$ and timing $\sigma_T(\zeta)$ jitter [see \eqn{sigma}] for \figs{figpdf}(a) and (b), respectively.
Because the characteristic functions of \eqn{cfFreq} and \eqn{cfTime} are even real functions, the probability density functions are also even functions.
\figs{figpdf} just plot for the positive frequency and timing jitter.
Comparing the frequency jitter of \fig{figpdf}(a) and the timing jitter of \fig{figpdf}(b), the frequency jitter has a distribution more close to the Gaussian distribution than that of timing jitter.

\begin{figure}
\centerline{
  \begin{tabular}{cc}
	 \includegraphics[width = 0.25 \textwidth]{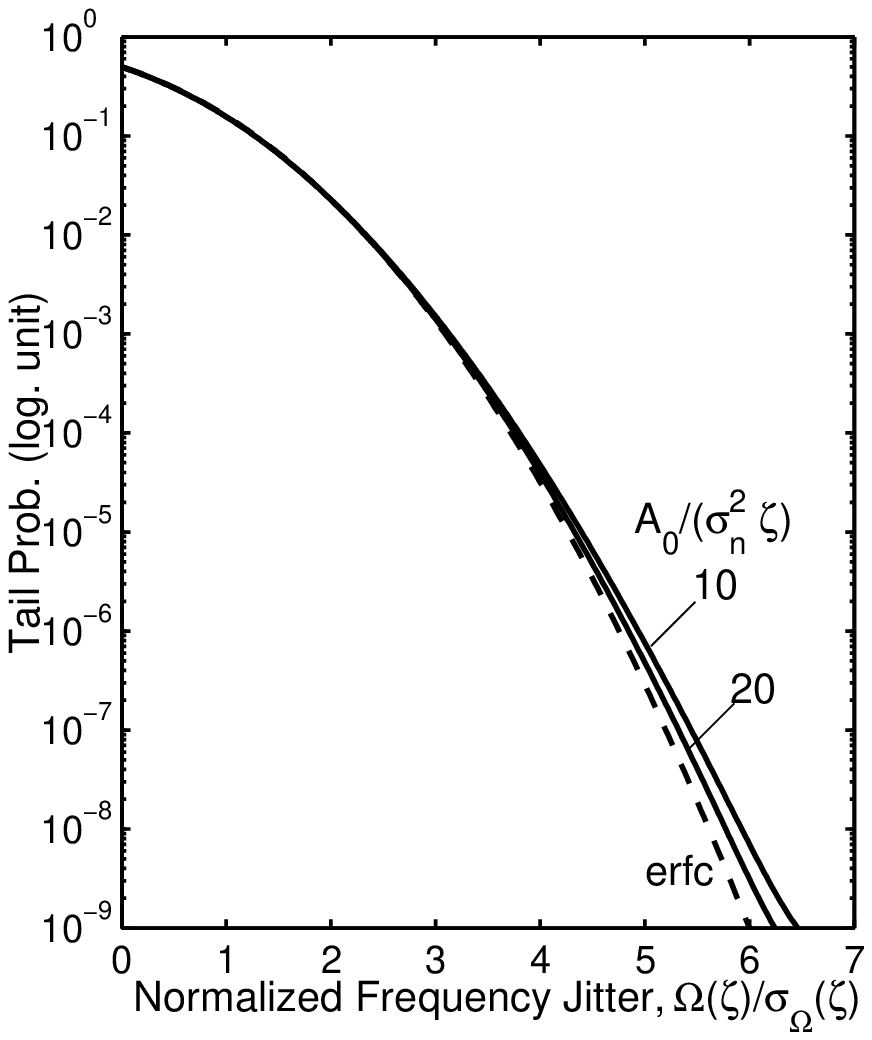} &  \includegraphics[width = 0.25 \textwidth]{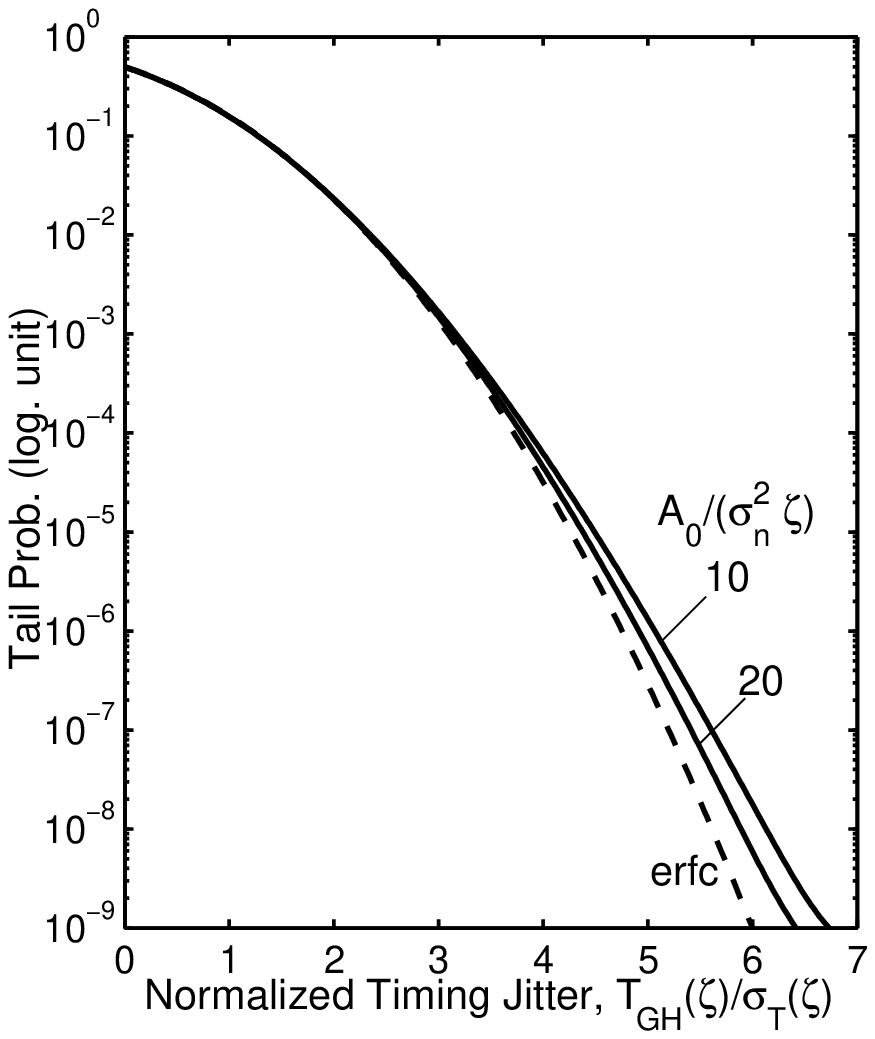} \\
	 (a) & (b) 
  \end{tabular}
}
\caption{The tail probabilities of (a) frequency and (b) timing jitter as compared with complementary error function.}
\label{figtail}
\end{figure}

\figs{figtail} plot the tail probabilities corresponding to the probability density functions of \figs{figpdf}. 
The tail probability is defined as $\int_x^\infty p(x) \ud x$ for a probability density function of $p(x)$.
The tail probability is compared to the complementary error function of $\frac{1}{2} \mbox{erfc}\left({x}/{\sqrt{2}}\right)$ that is the tail probability of the Gaussian distribution in \figs{figpdf}.
Comparing the tail probability of frequency jitter of \fig{figtail}(a) and the timing jitter of \fig{figtail}(b), the frequency jitter has a distribution more close to the Gaussian distribution than that of timing jitter.

From both \figs{figpdf} and \ref{figtail}, both frequency and timing jitters have a larger tail probability than that of the Gaussian distribution.
The non-Gaussian distribution leads to higher error probability than that of Gaussian distribution.

From both \figs{figpdf}(b) and \ref{figtail}(b), the distribution of Gordon-Haus timing jitter has a large different to Gaussian distribution at the tail.
Comparing the generating function of \eqn{cfC2a} and the Gaussian approximation of \eqn{cfApprox}, the non-Gaussian timing jitter distribution is given by the factor of $ \left[f_\lambda(\zeta)/ f_\lambda(0)\right]^{1/2} \exp(\lambda^2 \sigma^2_n A_0 \beta^2)$ that is obviously non-Gaussian.
Comparing the generating function of \eqn{cfC1a} and the Gaussian approximation of \eqn{cfApprox}, the non-Gaussian frequency jitter distribution is given by two factors: the third and higher-order powers of $\tanh(x) = x -x^3/3 + \cdots$ and the factor of $\mbox{sech}^{1/2}(\zeta \sigma_n \sqrt{\lambda}/2)$.

In conclusion, based on the first-order perturbation theory of soliton, both frequency and timing jitters are found to be non-Gaussian distributed.
Amplitude, frequency, and timing jitters are all Gaussian distributed if the equations from perturbation are linearized.
Without linearization, the noise projected into frequency and timing jitters are modulated by the amplitude jitter, leading to non-Gaussian distribution.
The timing jitter has larger discrepancy to Gaussian distribution than that of frequency jitter.

\bibliographystyle{apsrev}
\bibliography{../bib/kerr}%

\end{document}